\documentclass[aps,prl,twocolumn,groupedaddress,showpacs]{revtex4}
\usepackage{graphicx}
\usepackage{amsmath}
\usepackage{amssymb}
\usepackage{mathrsfs}
\usepackage{subeqnarray}

\begin{document}

\title{Morphogenesis of growing soft tissues\\}
\author{Julien Dervaux \\ Martine Ben Amar}
\address{Laboratoire de Physique Statistique, Ecole Normale Sup\'{e}rieure, 24 rue Lhomond, 75231 Paris Cedex 05}


\date{\today}

\begin{abstract}

Recently, much attention has been given to a noteworthy property of some soft tissues: their ability to grow. Many attempts have been made to model this behaviour in biology, chemistry and physics. Using the theory of finite elasticity, Rodriguez has postulated a multiplicative decomposition of the geometric deformation gradient into a growth-induced part and an elastic one needed to ensure compatibility of the body. In order to fully explore the consequences of this hypothesis, the equations describing thin elastic objects under finite growth are derived. Under appropriate scaling assumptions for the growth rates, the proposed model is of the F\"{o}ppl-von K\'{a}rm\'{a}n type.  As an illustration, the circumferential growth of a free hyperelastic disk is studied.

\end{abstract}

\maketitle

Biological tissues are conventionally classified into two categories: hard tissues (e.g. bones or teeth) and soft tissues (e.g. muscles, arteries, tendons, skin), depending on their mechanical properties. Soft tissues, which typically exhibit anisotropic, nonlinear, inhomogeneous behaviours, are often subject to large stresses and strains. The theory of finite elasticity therefore forms an appropriate framework to describe their properties \cite{fung1, shalak0, fung2}, in the absence of viscolelastic effects. Along these lines, much work has been done to establish constitutive relationships for specific biological materials such as the skin, blood vessels, lung, brain, liver and kidney \cite{fung2, miller}, although computing stresses and strains under applied external loads remains a difficult task.

Observation of biological tissues has revealed the existence of internal stresses, even in the absence of external loads. These \textit{residual stresses} are induced by growth \cite{shalak0} and affect the geometrical properties of tissues. Soft tissues may undergo \textit{volumetric growth}, that is a smooth volumetric resorption of the bulk material \cite{humphrey, cowin} depending on space, orientation and the state of stress within the body. Growth is a complex process involving biochemical and physical reactions at many different length- and time- scales, that occur through cell division, cell enlargement, secretion of extracellular matrix or accretion at surfaces. The removal of mass is referred to as \textit{atrophy} and occurs through cell death, cell shrinkage or resorption. The general idea is that the total deformation of the body can be due to both change of mass and elastic deformations \cite{hsu, cowin2, shalak3, entov, drozdov, stein}.

Before (resp. after) the deformation, the body is in the \textit{reference (resp. current) configuration} and the place of each material point is denoted by $\mathbf{X}$ (resp. $\mathbf{x}$). We define the \textit{geometric deformation tensor} by $\displaystyle \mathbf{F}= \frac{\partial \mathbf{x}}{\partial \mathbf{X}}$ to describe locally the \textit{overall} deformation process. In order to model the growth process, we follow Rodriguez \textit{et al} \cite{rodriguez} in making the following three assumptions: \textit{(i)} there exists a zero-stress reference state; \textit{(ii)} the geometric deformation gradient $\mathbf{F}$ admits a multiplicative decomposition of the form $\mathbf{F} = \mathbf{A} \mathbf{G}$ where $\mathbf{G}$ is a \textit{growth tensor} describing the change in mass and $\mathbf{A}$ an \textit{elastic tensor} characterizing the reorganization of the body needed to ensure compatibility (no overlap) and integrity (no cavitation) of the body; \textit{(iii)} the response function of the material depends only on the elastic part of the total deformation. Despite its simplicity, Rodriguez theory is yet to be investigated, because of the complexity of finite elasticity although inhomogeneous and anisotropic growth has been studied in details in some simple geometry \cite{benamar1, benamar2}. In order to explore the consequences of Rodriguez's model, we have written the equilibrium equations of a thin elastic body subject to growth-induced finite displacements. This reduction of dimensionality allows deep theoretical inspection and is relevant for many biological systems (e.g. leaves, skin). 

\begin{figure}[h!]
\begin{center}
	 \includegraphics[width=210pt,height=155pt]{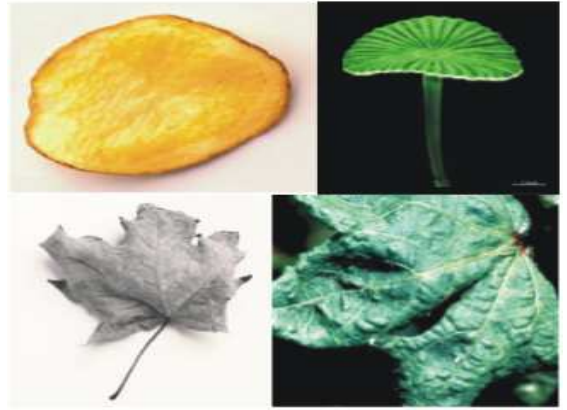}
        \caption{\footnotesize{A few examples of buckling in nature. (A) A potato chip adopts a saddle shape during frying. (B) \textit{Acetabularia schenckii}: a green algae.  (C) A dead leaf. (D) A leaf infected by the Cotton Leaf Crumple Virus (CLCrV), known to affect growth and induce curling and the appearance of blisters at the surface of the leaf.}}
\label{fig1}
\end{center}
\end{figure}

Under appropriate scaling assumptions, the resulting equations are found to be an extension of the well known F\"{o}ppl von K\'{a}rm\'{a}n (FvK) model, a powerful theory for buckling instabilities, that are widely diffused in nature (FIG. \ref{fig1}), but which is also able to explain complex post-buckling phenomena such as crumpling. Experimentally, it has been shown that growth may affect curvature in various systems. In growing gels, both homogeneous growth under constraints \cite{boudaoud} and free inhomogeneous growth \cite{sharon} have been investigated. Thermal expansion (FIG.1-A), as well as dessication, can also bend an elastic body and cause it to crumple as seen in dead leaves (FIG.1-C). In living tissues, viruses such as the Cotton Leaf Crumple Virus (CLCrV) modify the growth process and infected plants exhibit curled or crumpled leaves (FIG.1-D) but buckling can also occur during normal development. Some mushrooms' or algae's caps (FIG.1-B) may undergo symmetry breaking, and adopt an oscillatory or cup shape. At the cellular level, a new milestone was reached with the discovery of the \textit{CINCINNATA} gene whose local expression affects growth and curvatures of the \textit{Antirrhinum} (snapdragon) leaf \cite{nath}. Complementary to the inhomogeneity of growth, anisotropy has been shown to be crucial in the generation of shape. Indeed "\textit{a key aspect of shape -petal asymmetry- in the petal lobe of \textrm{Antirrhinum} depends on the direction of growth rather than regional differences in growth rate}" \cite{lagand}. To investigate the effects of anisotropy, for which our formalism is well suited, we study the problem of a free elastic disk subject to homogeneous anisotropic growth. 

\paragraph{The model.} 

Since biological soft tissues have a high volume fraction of water they are \textit{elastically incompressible}; in our notation $\det \mathbf{A}=1$. Furthermore, we assume isotropy of the material for simplicity and we define a strain energy function  $ {\cal W}  = \sum_{r,s=0}^\infty c_{rs} ( {\cal I}_1-3)^r( {\cal I}_2-3)^s $ where ${\cal I}_1 $ and ${\cal I}_2$ are the principal invariants of the tensor $ \mathbf{A}^{t} \mathbf{A}$. Note that any of the common constitutive relationships can be described by a series of this form \cite{ogden}.  After the deformation process, the sheet, of lateral size $L$ and thickness $H$, is described by the displacement field: $ \mathbf{u}= \mathbf{x}- \mathbf{X}$ and we define $\zeta(X,Y) = u_Z(X,Y,0)$ the displacement of the middle surface. When the growth rates $g_{ij}=G_{ij}-\delta_{ij}$ are in the range $(H/L)^2 \lesssim g_{ij} \ll 1$, the scaling of the induced strains falls inside the domain of validity of the FvK model. Since $H \ll L$, we also apply the membrane assumption that states $\sigma_{XZ}=\sigma_{YZ}=\sigma_{ZZ}=0$. Under these assumptions, after a rather long calculation, we derive two main conclusions:\textit{(i)} The in-plane Green tensor, defined as $\mathbf{E}=\frac{1}{2} \left( \mathbf{A}^t \mathbf{A} - \mathbf{I} \right)$ can be \textit{additively decomposed} into geometric and growth-induced part: $\mathbf{E}=\frac{1}{2} \left( \mathbf{F}^t \mathbf{F} - \mathbf{G}^t\mathbf{G} \right)$. \textit{(ii)} All materials behave according to the constitutive equation $\boldsymbol{\sigma} = \frac{2 Y}{3} \left( \mathbf{E} - p \mathbf{I} \right)$, $p$ being the hydrostatic pressure associated with the incompressibility constraint and $Y = 6(c_{01}+c_{10})$ the instantaneous Young modulus. Thus \textit{all thin elastic samples, undergoing small (but finite) deflexions follow a generalised Hooke's law whatever the constitutive relationship is}, as previously noted in \cite{erbay}. The pressure is given by the assumptions $\sigma_{ZZ}=0$, which implies $p=E_{ZZ}$. Moreover, $E_{ZZ}$ can be expressed in terms of the in-plane components of the Green tensor, since the incompressibility constraint yields $\mbox{Tr}\mathbf{E}=0$.  Once these results are established, we derive the equilibrium equations using the principle of minimal energy. They can be written in terms of the off-plane displacement and stresses:

\begin{subeqnarray} \label{eq1}
 \slabel{eq1a}
D \left( \Delta^2 \zeta -\Delta \phi \right) - H \frac{\partial}{\partial X_\beta} \left( \sigma_{\alpha \beta} \frac{\partial \zeta}{\partial X_\alpha}  \right)  & = & P, \\
 \slabel{eq1b}
\frac{\partial \sigma_{\alpha \beta}}{\partial X_\beta}  & = & 0,
\end{subeqnarray}

\noindent where Einstein summation convention is used, indices run from $1$ to $2$, $D=YH^3/9$ is the bending rigidity of the plate and $\phi$ is a source of mean curvature linked to the growth tensor via $\phi = \mbox{\bf{Div}} \left(\mathbf{G}\mathbf{G}^{t}\right)  \cdot \mathbf{e}_Z$. Using the Airy potential $\chi$:

\begin{subeqnarray} \label{eq2}
 \slabel{eq2a}
D \left( \Delta^2 \zeta -\Delta \phi \right) - 2 H [ \chi, \zeta ] & = & P, \\
 \slabel{eq2b}
\Delta^2 \chi + E \left( [\zeta, \zeta] -\psi \right)&=& 0,
\end{subeqnarray}

\noindent where the $[.,.]$ operator is defined in \cite{landau} and the function $\psi$ appearing in (\ref{eq2}) is a source of Gaussian curvature explicited in \cite{note1}.

The sets of equilibrium equations (\ref{eq1}) and (\ref{eq2}) are a generalization of the well known FvK theory of thin plates, to which they reduce in absence of growth, i.e, $\mathbf{G}= \mathbf{I}$. First consider the case where there exists a bijection $\overline{\mathbf{u}}$ such that $\mathbf{G} =  \mbox{\bf{Div}}\overline{\mathbf{u}}$. Then, provided that $\overline{\mathbf{u}}$ is consistent with the boundary conditions, then $\overline{\mathbf{u}}$ is a solution of the system of equations with zero elastic energy. In this case, the growth process is compatible (with itself and the boundary conditions) and there is no need for an elastic process, the shape being entirely determined by growth. Needless to say, this case is atypical and in general growth is incompatible, so neither $\mathbf{G}$ nor $\mathbf{A}$ are gradients of a deformation field (a one-to-one mapping). Thus the grown "state" cannot be physically achieved and is not referred to as a configuration. For large deformations however ($\zeta \gg H$), the problem can be simplified. Indeed the bending term $D \left( \Delta^2 \zeta -\Delta \phi \right)$ can be neglected and a solution that cancels the in-plane stresses is a solution of $[\zeta,\zeta] = \psi$, called a Monge-Amp\`{e}re equation. Once this equation is solved, the parameters appearing in this solution can be selected through minimization of the bending energy. For moderate deflections, i.e. $\zeta \sim H$, both bending and stretching terms are of the same order and the solution of zero energy is a surface with prescribed curvatures, which does not always exist; for example there are no surfaces that has positive Gaussian curvature and zero mean curvature. It is known that inhomogeneous growth can lead to sophisticated surface geometries \cite{sharon}, so we focus the research on anisotropic growth, which has been much less studied.

\paragraph{The free disk.} 

Consider a disc, of initial radius $R_i$, subject to anisotropic homogeneous growth, with free boundaries and no external loading.  Referring to a cylindrical system of coordinates $(R, \Theta, Z)$, the growth tensor is diagonal and homogeneous:

\begin{displaymath}
\mathbf{G} = \left(
\begin{array}{ccc}
1+g_1 & 0 & 0 \\
0 & 1 + g_2 &0 \\
0 & 0 & 1
\end{array}
\right)
\end{displaymath}

\noindent neglecting the thickening of the plate. If $g_1$ and $g_2$, respectively the radial and circumferential components of the growth process, are equal, then growth is homogeneous and isotropic and no residual stress appears: the disk remains flat. The relevant control parameter is $k = g_2-g_1$. The first case to consider is for $k \ll \frac{H^2}{R_i^2}$ that induces an off-plane displacement $\zeta$ much smaller than $H$ and is thus outside the scope of the present theory. When $k$ is of order $ \frac{H^2}{R_i^2}$, which leads to $\zeta \sim H$, all the contributions are of the same order and a linear stability analysis is performed. We look for a solution in which the in-plane fields (displacements $U_R, U_{\Theta}$ and stresses $\sigma_{RR}, \sigma_{R\Theta}$ and $\sigma_{\Theta \Theta}$) are independent of $\Theta$. The off-plane displacement, however, can depend on $\Theta$. Since the disk is free, the boundary conditions imply that there is no tension or torque at the free edge and reads $\sigma_{RR}(R_i) = \sigma_{R\Theta}(R_i) = 0$. The only convergent solution of (\ref{eq1b}), that fulfills these boundary conditions, is $\sigma_{RR} = \sigma_{R\Theta} = 0$ leading to $U_{R}(R)=\frac{2 R}{3}(g_2/2+g_1)$, $U_{\Theta}(R) = 0$ and a non-zero hoop stress $\sigma_{\Theta \Theta} = \frac{-2 k Y}{3}$.  Assuming a solution with discrete axial symmetry: $\zeta(\rho,\Theta)=\xi(\rho) \cos (m\Theta)$, (with $\rho=R/R_i $) equation (\ref{eq1a}) reduces to

\begin{eqnarray}
\label{eq:nonsymmetricsolution}
\nonumber
\xi^{(4)} + \frac{2}{\rho} \xi^{(3)} -  \frac{1 + 2 m^2}{\rho^2} \xi^{(2)}  & + \\
\frac{1 + 2 m^2 + \alpha \rho^2}{\rho^3} \xi^{(1)} + \frac{m^4 - 4m^2 -m^2 \alpha \rho^2}{\rho^4} \xi & = 0
\end{eqnarray}

\noindent where $\alpha= \frac{6 k R^2}{H^2}$ is a control parameter and $\xi^{(i)}$ is the i\textit{th} derivative of $\xi$ with respect to $\rho$. At the free edge ($\rho=1$) the zero-torque conditions are not affected by the growth process and are described in \cite{landau}. To avoid singularities at $\rho=0$, we impose $\xi(0)=0$ and $\xi'(0)=0$. These boundary conditions, together with equation (\ref{eq:nonsymmetricsolution}), form an eigenvalue problem for the threshold $\alpha$. Using Frobenius' method \cite{ince}, we find the four eigenfunctions corresponding to each $m$. The most unstable mode, occurring when growth is mainly circumferential ($\alpha > 0$), is characterised by $m=2$ -a saddle shape- with a threshold value of $\alpha = 3.08$. An axially symmetric solution, i.e $m=0$, appears when radial growth dominates ($\alpha < 0$), at the threshold value $\alpha = -7.82$. 
\begin{figure}[h!]
\begin{center}
				\includegraphics[width=240pt,height=180pt]{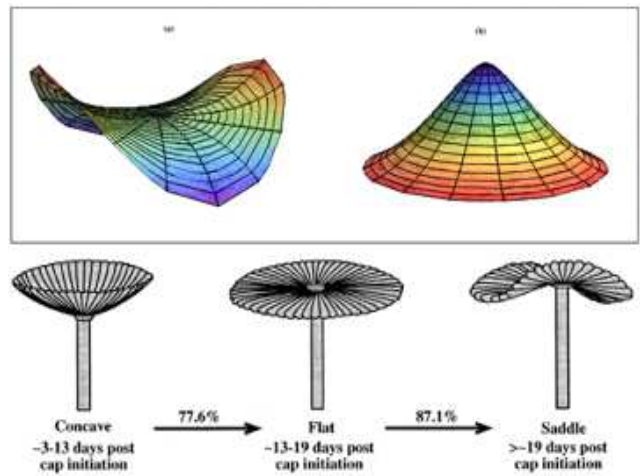}
         \caption{\footnotesize{Top: The two first destabilized modes. (a) On the left $k>0$, the disc adopts a saddle shape, with $m=2$, at the threshold $\alpha = 3.08$. (b) On the right $k<0$ and the disc adopts an axially symmetric shape characterized by $m=0$, at the threshold $\alpha = -7.82$. Bottom: shape changes in the \textit{Acetabularia} algae, the figures indicate the fraction of algae that undergo the shape transition from an initial population of 85 plants, picture drawn from \cite{serikawa}.}}
\label{fig2}
\end{center}
\end{figure}
This simple model explains surprisingly well the changes of cap shape that the algae \textit{Acetabularia acetabulum} undergoes during its development. Experiments performed in \cite{serikawa} show that radial growth occurs in the earliest stage of the development, which leads to a symmetric conical shape. At later stage however circumferential growth predominates to produces the saddle shape (see FIG.2).

We now consider large deformations: $1 \gg k \gg \frac{H^2}{R_i^2}$, for which $\zeta \ll H$. Since the stretching contribution is much bigger than the bending energy, we first solve the Monge-Amp\`{e}re equation $[\zeta,\zeta]=\psi$ in which $\psi$ is given by $\psi= k \delta(\rho) / \rho$ in our case. The general solution is a cone that has zero Gaussian curvature except at the tip of the cone where the effect of bending becomes important and which would require a more precise treatment \cite{benamar0,mahadevan}. We only focus on the outer solution. The equation of the cone is simply $\zeta(\rho,\Theta) = \rho g(\Theta)$. Using this expression, the condition that the Airy potential vanishes everywhere (so (\ref{eq2a}) and (\ref{eq2b}) are satisfied) gives for the in-plane displacement field:

\begin{equation}
U_\rho=-\frac{\rho}{2} g(\Theta)^2 \quad U_\Theta = \frac{\rho}{2} \int^\Theta \left(g(\Theta)^2-g'(\Theta)^2 + 2 k \right) {\mbox d}\Theta
\end{equation}

Periodicity in the orthoradial displacement implies $U_\Theta(\rho,0)=U_\Theta(\rho,2 \pi)$. Let $g(\Theta)$ be represented by its Fourier series: $ g(\Theta)= \sum_{n=0}^{\infty}\left(a_n \mbox{e}^{\mbox{i} n \Theta} + a^{\star}_n \mbox{e}^{-\mbox{i} n \Theta} \right)$. Periodicity condition leads to:

\begin{figure}[h!]
\begin{center}
\label{fig3}
        \includegraphics[width=240pt,height=190pt]{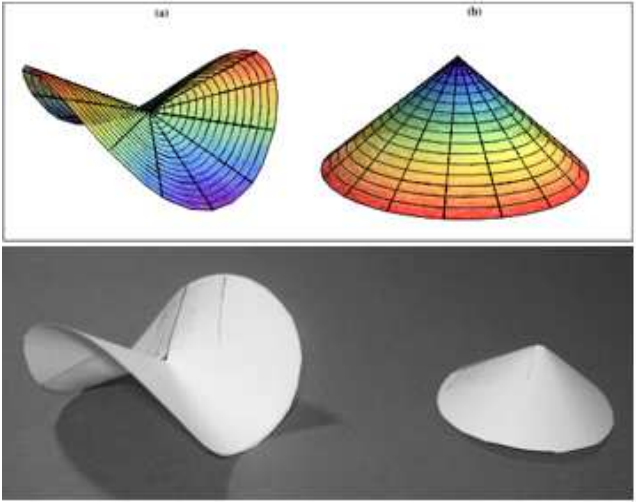}
        \caption{\footnotesize{Top: The two modes of minimal energy, far from the tip. (a) $k>0$, the shape is a cone with two oscillations (b) $k<0$ the cone of revolution is a solution. Bottom: The resultant shapes built out of paper ($k=0.25$) are in agreement with the prediction.}}
\end{center}
\end{figure}

\begin{equation}
\label{eq3}
2 \sum_{n=1}^{\infty}\left(a_n a^{\star}_n (n^2-1) \right) = (a_0 + a^{\star}_0)^2 + 2 k
\end{equation}

A cone of revolution (for which $a_0$ is the only non-zero coefficient of the Fourier series) can satisfy the periodicity condition only if $k<0$; that is when radial growth dominates. Infinitely many solutions satisfy the condition (\ref{eq3}) but the selected shape must have minimal bending energy \cite{benamar0,mahadevan,mahadevan1}. The bending contribution reads:

\begin{equation}
\label{eq4}
{\cal E}_b \propto \left[ 2 \sum_{n=1}^{\infty}\left(a_n a^{\star}_n (n^2-1)^2 \right) + (a_0 + a^{\star}_0)^2 \right]
\end{equation}
 
Calculating the sum in (\ref{eq4}) we find the solutions of minimal outer bending energy:

\begin{subeqnarray}
k<0 \qquad \zeta(R,\Theta) & = & R \sqrt{2 k} \\
k>0 \qquad \zeta(R,\Theta) & = & R \sqrt{\frac{4 k}{3}} \cos{2 \Theta}
\end{subeqnarray}

For large deformations, those predictions can be easily checked by constructing a cone from a disc of paper in which a sector defined by two radii is withdrawn and then either replaced by a bigger one or just glued to close it. This simple demonstration illustrates the fact that singularities can arise from growth as observed in dead leaves or in the leaves infected by the CLCrV.

\paragraph{Conclusion} Using the formalism introduced by Rodriguez \textit{et al}, we have developed a theory describing the behaviour of thin elastic bodies subject to growth. By expliciting the sheet's small thickness, we showed all materials behave according to a generalized Hooke's law and the equilibrium equations generalize the FvK equations accounting for growth. This extension describes a broad range of physical phenomena involving mass reorganization, from biological growth to thermal dilatation, as well as dessication. The treatment presented in this letter makes the inclusion of anisotropic effects and spatial inhomogeneities of the growth process easy. We have shown that anisotropic growth induces rich structures like curling and crumpling.

\end{document}